\documentstyle[psfig,subfigure,lscape]{mn}

\begin{document}
\newcommand{\goodgap}{%
 \hspace{\subfigtopskip}%
 \hspace{\subfigbottomskip}}

\title[A census of metals at high and low redshift]{A census of metals
at high and low 
redshift and the connection between submillimetre sources and spheroid
formation.}
\author[L. Dunne, S. A. Eales, M. G. Edmunds]
{Loretta Dunne$^{1}$, Stephen A. Eales$^1$, M. G. Edmunds$^1$\\
$^1$Department of Physics and Astronomy, University of Wales Cardiff,
PO Box 913, Cardiff, CF2 3YB\\ }

\maketitle

\begin{abstract}
Deep surveys in many wavebands have shown that the rate at which stars
were forming was at least a factor of 10 higher at redshifts $>1$ than
today. Heavy elements (`metals') are produced by stars, and the star
formation history deduced by these surveys implies that a significant
fraction of all metals in the universe today should already exist at
$z \sim 2-3$. However, only 10\% of the total metals expected to exist
at this redshift have so far been accounted for (in Damped Lyman Alpha
absorbers and the Lyman forest). In this paper, we use the results of
submillimetre surveys of the local and high redshift universe to show that
there was much more dust in galaxies in the past. We find that a large
proportion of the missing metals are traced by this dust, bringing the
metals implied from the star formation history and observations into
agreement. We also show that the observed distribution of dust masses
at high redshift can be reproduced remarkably well by a simple model
for the evolution of dust in spheroids, suggesting that the
descendants of the dusty galaxies found in deep submm surveys are the
relatively dust-free spiral bulges and ellipticals in the universe
today.
\end{abstract}

\begin{keywords}
galaxies:evolution -- galaxies:ISM -- galaxies:luminosity function,
mass function -- submillimetre -- (ISM):dust, extinction
\end{keywords}

\section{Introduction}
\begin{figure}
\psfig{file=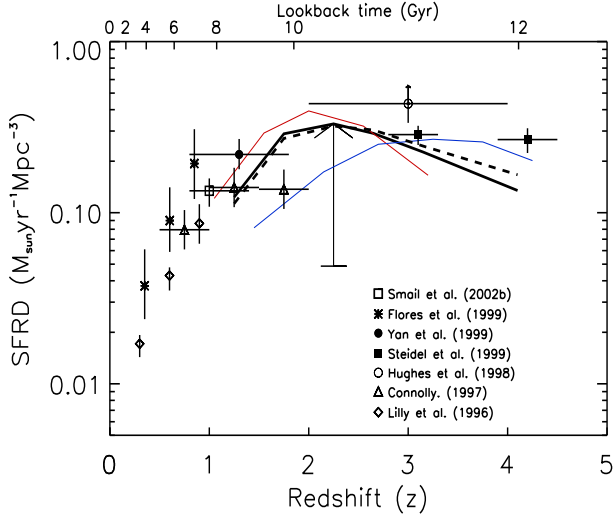,width=8cm,height=7cm}
\caption{\label{sfrdF} The current best estimate of the star formation
history of the universe, taken from Ivison et al. 2002. The points
represent extinction corrected optical/UV, radio and submm estimates
while the lines are the latest submm estimates based on the 8mJy
survey (Scott et al. 2002) using different redshift estimators. The
arrow represents the extrapolation of the 8 mJy population to 1 mJy
(see Ivison et al. 2002). All
estimates are in reasonable agreement, leading to a star formation
history which is relatively constant between $z=1-4$, declining by a
around a factor of 10 from $z=1$ to today.} 
\end{figure}

The star formation history (SFH) of the universe deduced from
optical/UV surveys (Lilly et al. 1996; Madau et al. 1996, 1998) has
recently been revised to account for dust extinction (Steidel et
al. 1999), largely motivated by the detection of a population of dusty
star-forming objects in deep submm surveys with SCUBA (Hughes et al. 1998;
Eales et al. 1999; Smail, Ivison \& Blain 1997; Scott et
al. 2002). The current best estimate of the SFH is that the global
star-formation rate in the universe was fairly constant over the
redshift interval $1<z<4$, falling by a factor $>10$ to the present
epoch, see Fig.~\ref{sfrdF}. (Steidel et al. 1999; Ivison et
al. 2002). A simple integration of this history implies that $\sim
25$\% of all stars and metals today should have formed by a redshift
of $\sim 2.5$.\footnote{In a flat cosmology.} However, the metal
content of objects believed to contain most of the baryonic matter at
high-z -- principally the Lyman alpha forest and damped Lyman alpha
absorption systems (DLAs)-- is only 2.5\% of the total amount of
metals in the universe today (Pettini 1999; Pagel 2002). Thus 90\% of
metals expected to exist at $z=2-3$ have not yet been seen or
recognised -- the `missing metals'.

The submm surveys revealed a population of extremely dusty and
obscured objects with very high bolometric luminosities, probably
residing at high redshifts, $z>2$ (Dunlop 2001a; Ivison et
al. 2002). If the dust is heated by stars, the star formation rates
implied by the submm fluxes are $\sim 100-1000
\,\rm{M_{\odot}\, yr^{-1}}$, high enough to produce an entire galaxy
worth of stars in $\sim 1$ Gyr. Since the properties of low-redshift
ellipticals suggest that most of their stars formed quickly at high
redshift (Bower et al. 1992; Renzini 1997; Jimenez et al. 1999),
several authors have argued that the SCUBA sources may be elliptical
galaxies in the process of formation (Lilly et al. 1999; Eales et
al. 2000; Dunlop 2001b; Granato et al. 2001). It has been claimed (Cen
\& Ostriker 1999) that the regions of greatest over-density in the
universe at any epoch are the most metal enriched, and that these
regions will have reached a given metallicity much earlier than lower density
environments. This argument suggests that SCUBA galaxies should be a
large reservoir for metals at high redshift, since if they are
proto-ellipticals they should have formed from the highest density peaks at
high redshift when there has been less time for pollution of the low
density inter-galactic medium (IGM).

In this paper, we wish to investigate the connection between the
`missing metals', the formation of the spheroids and the dusty sources
seen by SCUBA. However, instead of using the bolometric luminosities
of the submm sources to infer their physical properties, we interpret
the results of the deep submm surveys and our submm survey of the
nearby universe in a completely different way, by carrying out the
first investigation into the cosmic evolution of the dust mass in
galaxies. This provides a new way of looking at the chemical evolution
of the universe, as there is evidence that the fraction of ISM metals
which are bound up in dust grains is $\sim 40$ percent, a value which
seems remarkably constant from galaxy to galaxy and which also appears
to apply at
higher redshifts (Issa, MaClaren \& Wolfendale 1990; Edmunds 2001;
Pei, Fall
\& Hauser 1999; James et al. 2002). We will determine
the dust content of the low and high-z universe by constructing dust
mass functions, i.e. the space densities of galaxies as a function of
dust mass. At low-z, we will use the results of the SCUBA Local
Universe Galaxy Survey (Dunne et al. 2000), the first attempt at an
unbiased submm survey of the local universe. For the high redshift
dust mass function, we will use submm data from the deep SCUBA submm
surveys. In Section 4 we will use these dust mass functions, together
with a simple chemical evolution model, to estimate the co-moving
density of metals and baryons associated with the ISM of galaxies at
low and high redshift. Finally, we will produce a simple prediction
for the dust mass function of the spheroids during the epoch in which they
had their maximum dust masses, and compare with our observational
estimate of the dust mass function of the SCUBA sources.

\section{Dust Masses from Submillimetre Fluxes}

Dust mass can be calculated from the submm flux according
to

\begin{equation}
M_{\rm{d}} = \rm {\frac{(1+z)\, D^2\, S_{obs}}{\kappa_d(\nu_{em})\,
B(\nu_{em},\, \langle T_d \rangle)}}  \label{mdustE}
\end{equation}

where $D$ is the distance for the assumed cosmology and $\rm{S_{obs}}$
is the observed flux. The dust mass opacity coefficient
$\kappa_{d(\nu_{\rm{em}})}$ normalises the submm emission for a given
dust mass and is currently measured only at FIR wavelengths and then
extrapolated to the submm with a $\lambda^{-\beta}$ dependency. We
have taken the average value from the following sources of
$\kappa_{d(125\mu m)} = 2.64 \pm 0.29\,\rm{m^2\, kg^{-1}}$, where the
$\pm 0.29$ is the error on the mean value (Hildebrand 1983; Draine
\& Lee 1984; Kr\"{u}gel, Steppe
\& Chini 1990; Sodroski et al. 1997; Bianchi, Davies \& Alton
1999). We use $\beta=2$ (for which there is compelling evidence from
our local survey and elsewhere, Dunne \& Eales 2001 and references
within) to extrapolate $\kappa_{d(125\mu m)}$ to the rest wavelength
of the submm emission ($\sim 240\mu$m for an 850$\mu$m source at
$z\sim 2.5$). This agrees with the recent estimation of
$\kappa_{d(850\mu m)} =0.07\pm 0.02\,\rm{m^2\, kg^{-1}}$ using a
different technique (James et al. 2002). 

We also require the value of $\langle T_d\rangle$, which is a
mass-weighted dust temperature. Submillimetre and ISO observations of
many nearby galaxies have shown that, when enough data is present, the
SED is best described by dust at more than one temperature (Haas et
al. 1998,2000; Frayer et al. 1999; Braine et al. 1997; Dumke et
al. 1997, Calzetti et al. 2000; Gu\'{e}lin et al. 1995; Sievers et
al. 1994; Neininger et al. 1996; Alton et al. 1998, 2001; Papadopoulos
\& Seaquist 1999; Reach et al. 1995; Trewhella et al. 2000). For our
local SCUBA survey of bright IRAS galaxies, we found that in addition
to the typical `IRAS-detected' dust at $35-45$ K, a colder dust
component at $17-25$ K was required in all cases (Dunne \& Eales
2001), even for highly luminous galaxies such as Arp 220. Klaas et
al. (2001) also find that the SEDs of 41 local ULIRGs are best
represented by a combination of warm and cold dust. Following the
notation of the two-component SED model from Dunne \& Eales (2001),
the value which should enter Eqn.~\ref{mdustE} can be approximated by
\begin{equation}
\langle T_d \rangle = \frac{2(T_w - T_c)}{(1+N_c/N_w)^2} + T_c \label{mtdE}
\end{equation} 
where $T_c/T_w$ is the characteristic temperature of the cold/warm
dust, and $N_c/N_w$ is the ratio of mass in the cold/warm
components. For the local survey, the value of $\langle T_d\rangle$
was 21.3 K, with a standard error in the mean for the population of
0.5 K. This mass-weighted temperature should not be confused with the
dust temperature traditionally used to estimate bolometric luminosity,
which should be the temperature defining the peak in the
spectrum. Since we are not integrating the SED and have observations
at wavelengths longer than the peak, our estimation of the dust mass
from a single submm flux is not as sensitive to temperature as an
estimate of the bolometric luminosity would be. In local galaxies,
including ULIRGs, most dust is contained in the cold $\sim 20$ K
component, which means we do not need to worry about the location of
the peak in the SED; the temperature of the warm dust has a negligible
effect on the dust mass estimate, providing $N_c/N_w > 1$. For the
local objects, we use $\langle T_d\rangle$ for each galaxy in our
survey (Dunne \& Eales 2001).

The biggest uncertainty in our analysis is our lack of knowledge of
$\langle T_d\rangle$ in the high redshift SCUBA galaxies. As discussed
above, there is now a lot of evidence that most of the dust in low
redshift FIR sources, including ULIRGs, is at a relatively cold
temperature. It will not be possible to perform a similar analysis of
the temperature of dust in high redshift galaxies because of the lack
of instruments operating at wavelengths long enough to sample the
rest-frame emission from the cold dust (i.e. 850$\mu$m in the {\em
rest-frame\/}). It is difficult to determine the existence of a colder
dust component without many fluxes at wavelengths at and beyond the
peak in the SED. It is plausible that the mean temperature of dust in
galaxies was higher in the past, as star-formation rates and possibly
interstellar radiation fields were higher. However, it is also
possible that the mass-weighted dust temperatures in high-z SCUBA
galaxies might be as low, or even lower than in low-z ULIRGs. This is
because dust equilibrium temperatures are very sensitive to the
relative geometry of dust and heating sources, as well as optical
depth effects. If the clouds in which the stars form are very dense,
it is likely that few UV/optical photons escape (we know this is the
case for the high redshift SCUBA galaxies as a whole), and so dust
outside these regions will see a much weaker radiation field and hence
have a lower temperature. There is also evidence (Frayer 2001; Ivison
et al. 2001,2002; Isaak et al. 2002; Lutz et al. 2001, Eales et
al. 2000) that the physical sizes of the star forming regions in some
submm luminous objects at high redshift are larger than in local
ULIRGs. Local ULIRGs have very compact regions of star formation ($<1$
kpc) while the radio, mm and CO sizes of the high-z submm sources have
been estimated to be as large as 10--30 kpc in some cases. This could
be due to the high-z objects having much greater gas fractions than
local ULIRGs. Mihos \& Hernquist (1996) have argued that it is the
presence of large stellar bulges in local ULIRGs which provide enough
stability to allow the gas to reach very high central concentrations
before forming stars. Galaxies at high redshift may have greater gas
fractions and small or non-existent stellar bulges compared to
galaxies today. This means that the star formation induced by an
interaction or perturbation at high redshift could be much more
widespread, as the gas will collapse to form stars before it inflows
to the centre of the remnant (Mihos 1999; Tissera et al. 2002). A
given star formation rate which is spread over a larger volume will
result in a lower intensity radiation field, possibly leading to lower
dust temperatures. Two sources at $z=0.5-1$ from the 170$\mu$m FIRBACK
survey have also been followed up with SCUBA and appear to have much
colder dust for their luminosity than local objects selected by IRAS
(Chapman et al. 2002). Such sources may be more representative of
objects selected in deep submm surveys, as selection at longer
wavelengths is less biased toward objects dominated by warmer
dust. The models of Pei, Fall \& Hauser (1999) however, predict that
there should have been a modest amount of positive evolution in the
dust equilibrium temperature between $z=0$ and $z=2-3$. In the absence
of any direct evidence on the amount of cold dust in the high-z SCUBA
galaxies, we have assumed the equilibrium temperature predicted for
dust sources at $z\sim 2-3$ by Pei, Fall \& Hauser (1999) of 25 K. If
temperatures are more akin to local galaxies (20 K) then the high-z
dust masses we present will increase by a factor $\sim 2$, and more if the
equilibrium temperatures at high redshift are actually lower than in
local ULIRGs.

\section{Dust Mass Functions at Low and High Redshift}

We estimate the space density of galaxies as a function of dust mass
(the dust mass function -- DMF) using standard accessible volume
techniques, such that
\begin{equation}
\Phi(M)\Delta M = \Sigma_i \frac{1}{V_i} \label{dmfE}
\end{equation}
where $V_i$ is the accessible volume in the parent sample of the $i$th
source and the sum is over all objects in the range $\rm{M
\rightarrow M+\Delta M}$. For the local galaxies $V_i$ is their
accessible volume at 60$\mu$m, which was the wavelength of selection of
their parent sample, the IRAS BGS (Soifer et al. 1989). The local dust
mass function (DMF) we will use here is described in detail in Dunne
\& Eales (2001).

\begin{figure}
\psfig{file=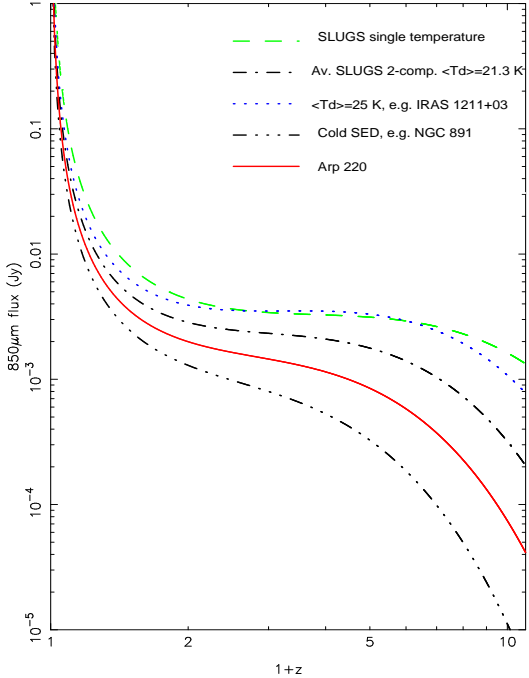,height=9cm,width=7cm}
\caption{\label{fluxzF} The 850$\mu$m flux density for a source with a
given dust mass in an $\Omega_{m}=0.3\,
,\, \Omega_{\Lambda}=0.7$ cosmology. Various SEDs are shown: SLUGS
single temperature is the average single temperature fit to the local
survey data ($\beta =1.3$, $T_d = 36$ K ) Dunne et al. (2000); average
SLUGS 2-component fit provides the mass-weighted temperature of 21.3
K, $[T_w=42$ K, $T_c=21$ K, $N_c/N_w =10]$; warmer SED with
$\langle T_d \rangle = 25$ K as assumed for the high-z sources, 
$[T_w =45$ K, $T_c=25$ K, $N_c/N_w = 13]$, similar to that fitted to
ULIRG IRAS 1211+03 (Dunne \& Eales 2001); cold SED for more quiescent
galaxies like NGC 891 (Alton et al 1998), $[T_w = 30$ K, $T_c=15$
K, $N_c/N_w = 100]$; the 2-component SED for Arp 220 (Dunne \& Eales
2001) $[T_w = 48$ K, $T_c =18$ K, $N_c/N_w = 42]$. All 2-component
SEDs have $\beta=2$. The fluxes for the 2-component SEDs drop off
after $z=4-5$, most dramatically for those with more cold dust.}
\end{figure}

For the high redshift dust sources, we construct the DMF using submm
data from the CUDSS blank field submm survey (Eales et al. 2000; Webb
et al. 2002), and also the submm number counts at higher and lower
flux densities (Smail et al. 2002; Scott et al. 2002; Borys et
al. 2002). In order to allow for star forming objects which may be too
submm faint to appear in the blank field surveys, we use the
statistical submm analysis of starburst galaxies in the HDF by Peacock
et al. (2000). The characteristic shape of the thermal dust spectrum
means that in a flat cosmology, an object of a given dust mass will be
observed to have virtually the same 850$\mu$m flux-density over a wide
range of redshift ($1<z<10$). Therefore the redshift we assume for the
sources has very little effect on the derived dust masses, providing
that the objects are at $z>1$ (any objects known to have $z<1$ were excluded
from this analysis). Spectroscopic redshifts (Eales et al. 2000) and
estimates using the Dunne, Clements
\& Eales (2000) version of the radio-submm redshift indicator (Carilli
\& Yun 1999) were used where available, otherwise sources were
placed at $z=2.5$ for the purposes of calculating
$M_{\rm{d}}$. Figure~\ref{fluxzF} shows that due to the flat
flux-density redshift relation, a source visible at $z=1$ will still
be visible (in a flux limited sample) at $z=5-10$ (depending on the
exact shape of the spectrum); therefore $V_i$ for the submm sources
was simply taken to be the volume in the redshift interval
$z_{\rm{max}}-z_{\rm{min}}$. We fix $z_{\rm{min}}=1$ as there is
dramatic evolution in the submm population below this redshift, very
few submm sources being found at $z<1$ (Dunlop 2001a), while we take
$z_{\rm{max}}=5$. This is the redshift at which the 850$\mu$m flux
density of a source with an SED similar to Arp 220 would start to fall
off (Fig.~\ref{fluxzF}). The effects of using $z_{\rm{max}}=4$
or $z_{\rm{max}}=6$ are a modest $\pm 20$\% change in the space
density. We have treated the statistical detection of submm flux in
the HDF in a similar manner to the blank field sources, using $z=3$,
the average of the photometric redshifts, to estimate the dust
mass. The photometric redshifts show that the galaxies responsible for
the submm emission are found between $z=1-6$ and this redshift range
was used to calculate the space density. There were also 5 optical
sources in the HDF which are plausible $>3\,\sigma$ detections in the
submm, these are treated separately and we have used their photometric
redshifts to estimate the dust mass. 

\begin{figure*}
\psfig{file=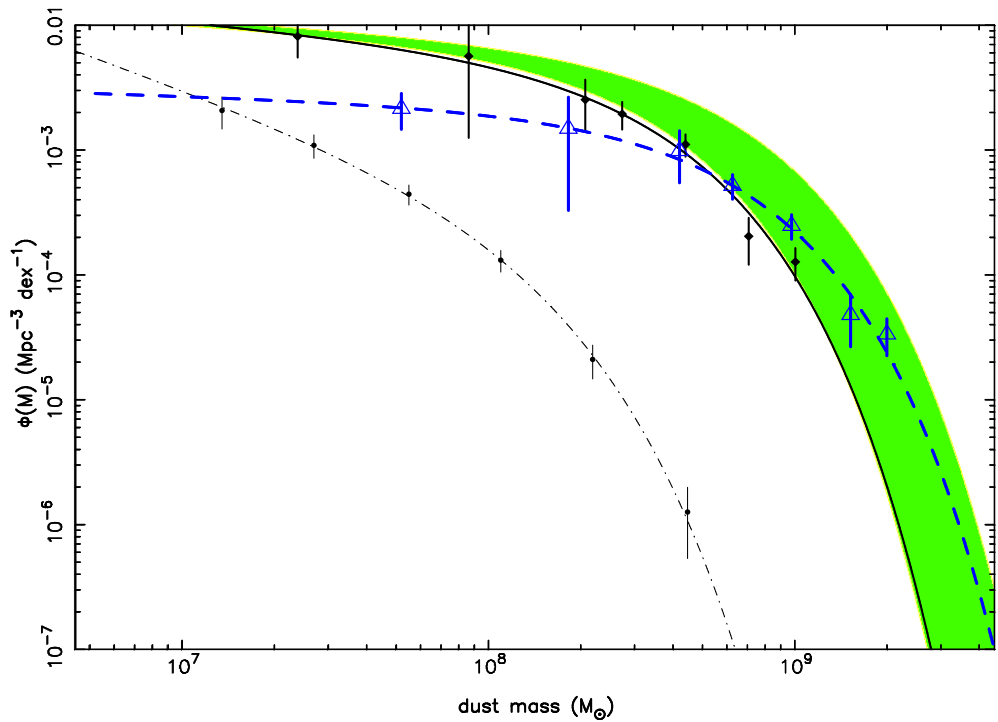,height=11cm,width=16cm,angle=-90}
\caption{\label{dmfF} Dust mass functions at high redshift
(diamonds and triangles) and locally (circles) along with best fitting
Schechter functions. The observations at high redshift match the
predictions for spheroids (shaded region) when their dust masses were
at a maximum. High-z points estimated using data from: CUDSS blank
field SCUBA survey (Eales et al. 2000; Webb et al. 2002) -- points
4,5,6; submm source counts at higher and lower fluxes (Smail et
al. 2002; Scott et al. 2002; Borys et al. 2002) -- points 2,7; submm
properties of optical starbursts in the Hubble Deep Field (HDF)
Peacock et al. (2000) -- points 1,3. Errors are simple Poisson
statistics. Blue dashed line and open triangles are for a
$\rm{\Omega_m=0.3,\,\Omega_{\Lambda}=0.7}$ cosmology, solid line and
points for $\Omega_m=1$. Shaded region indicates the maximal dust
evolution for local spheroids assuming $(M/L)_{\ast}= 4.5-8.3$ and a
closed box chemical evolution model. The space densities for the
high-z DMF change by $\sim \pm 20$\% if the volume is calculated using
$z_{\rm{max}}=4,6$ rather than 5. Schechter parameters for the high-z
DMF and $\rm{\Omega_m=1}$ are $\alpha = -1.22$,
$M^{\ast}_{\rm{d}}=2.7\times 10^8
\,\rm{M_{\odot}}$, $\phi^{\ast} = 2.4\times 10^{-3}\, \rm{Mpc^{-3}}$
($M^{\ast}_{\rm{d}}$ increases to $5.4\times 10^8 \,\rm{M_{\odot}}$ for
$\langle T_{d} \rangle = 20$ K) , and for the $\Lambda$ cosmology are
$\alpha = -1.08, \, M^{\ast}_{\rm{d}}=4.7\times10^8 \,\rm{M_{\odot}}, \,
\phi^{\ast} = 8.9 \times 10^{-4}\, \rm{Mpc^{-3}}$. The Schechter
parameters for the local DMF are $\alpha = -1.85$, $M^{\ast}_{\rm{d}}=
9.3\times 10^7 \,\rm{M_{\odot}}$, $\phi^{\ast} = 2.1\times 10^{-4}\,
\rm{Mpc^{-3}}$.}
\end{figure*}

This high-z dust mass function is shown in Fig.~\ref{dmfF} for two
cosmologies, $\rm{\Omega_m =1}$ (solid) and a $\Lambda$ model with
$\Omega_{\Lambda}=0.7,\,
\rm{\Omega_m =0.3}$ (blue dashed), along with the DMF from our local
survey. A value of $\rm{H_0 =75
\,km\, s^{-1} \, Mpc^{-1}}$ is used throughout. It is immediately
obvious that there has been substantial evolution in the dust mass
function, particularly at the highest dust masses.

\section{The Metal Content of the Universe}

\begin{figure*}
\subfigure[]{\psfig{file=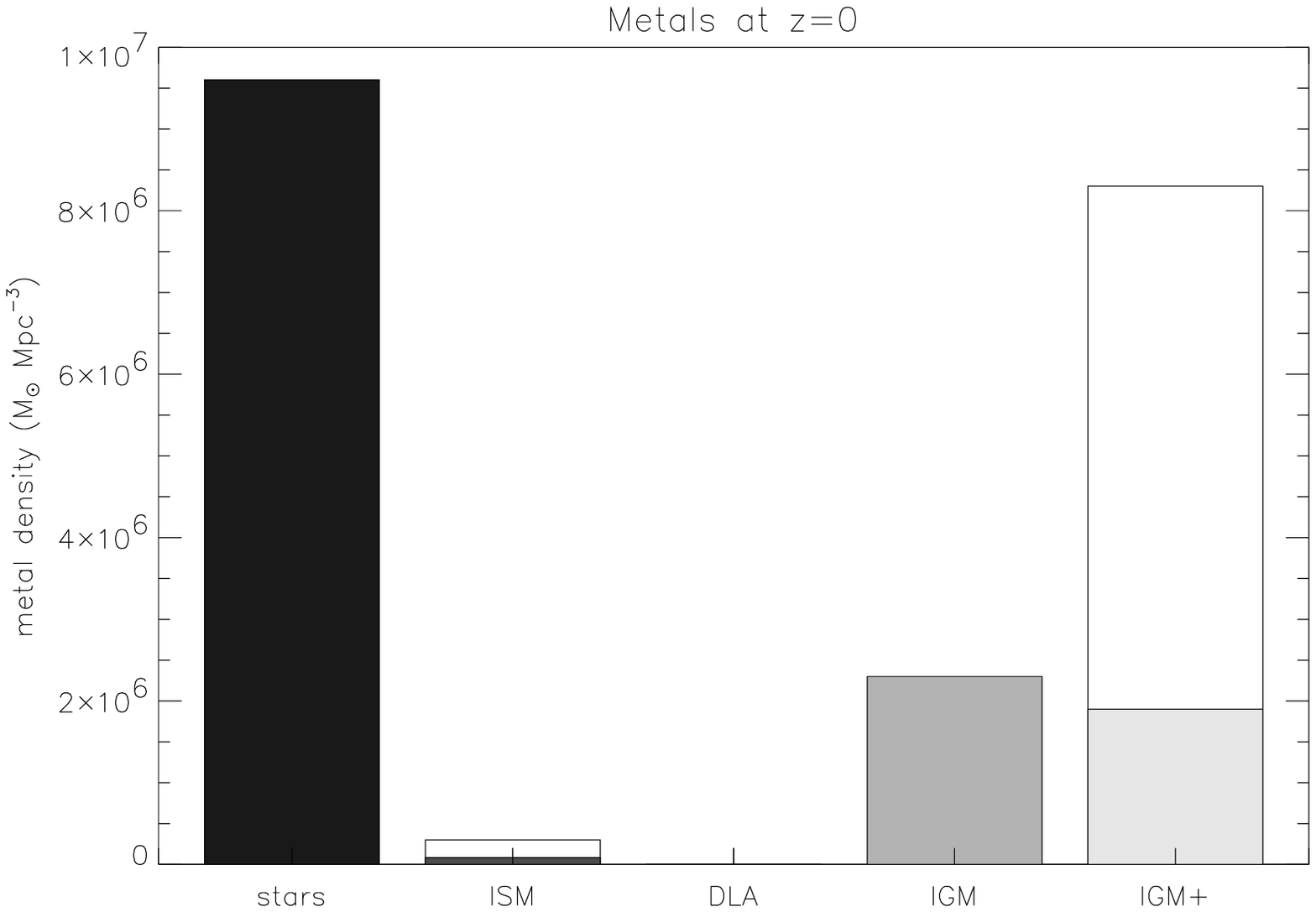,height=7.5cm,width=8.5cm}}\goodgap
\subfigure[]{\psfig{file=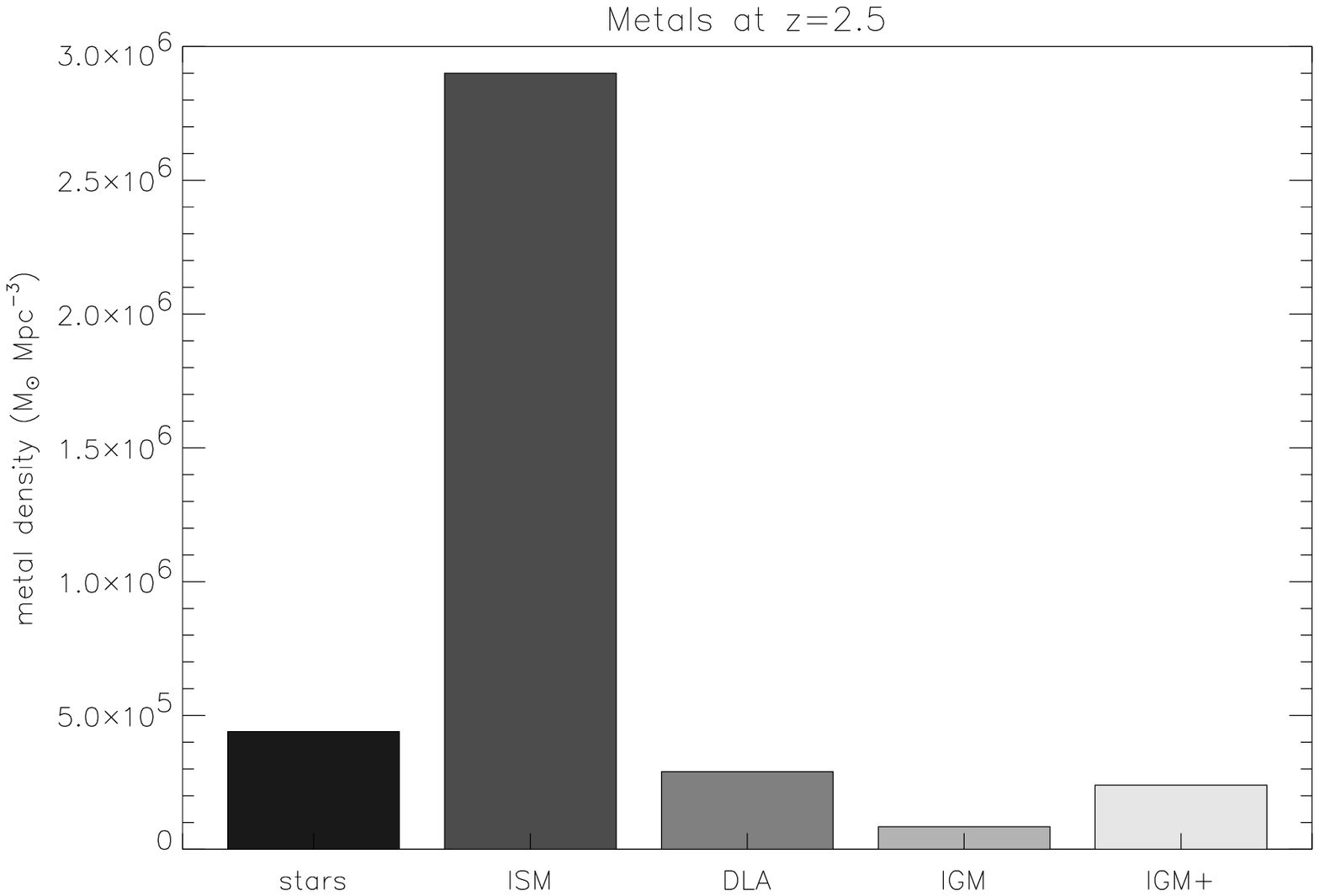,height=7.5cm,width=8.5cm}}\\
\subfigure[]{\psfig{file=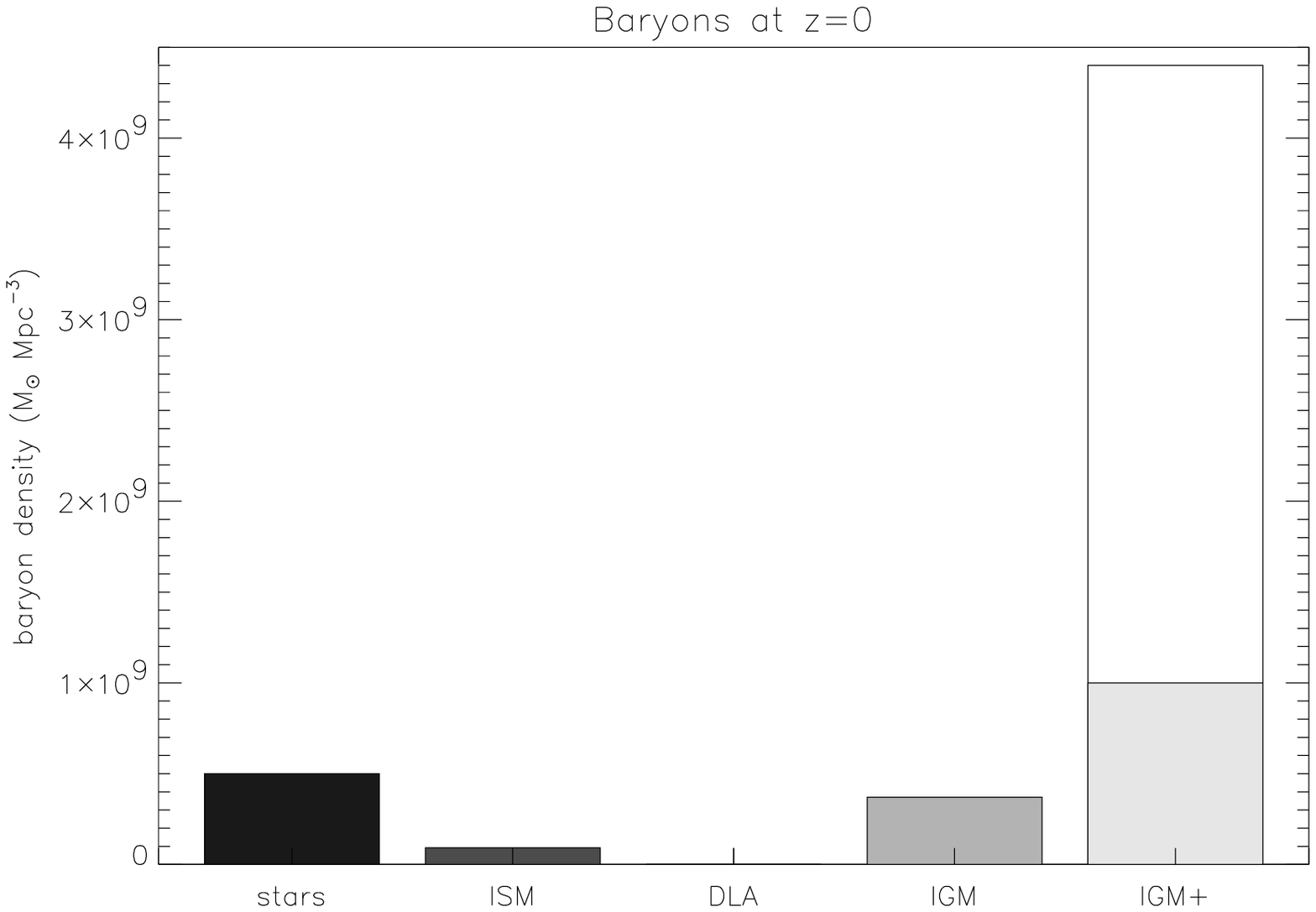,height=7.5cm,width=8.5cm}}\goodgap
\subfigure[]{\psfig{file=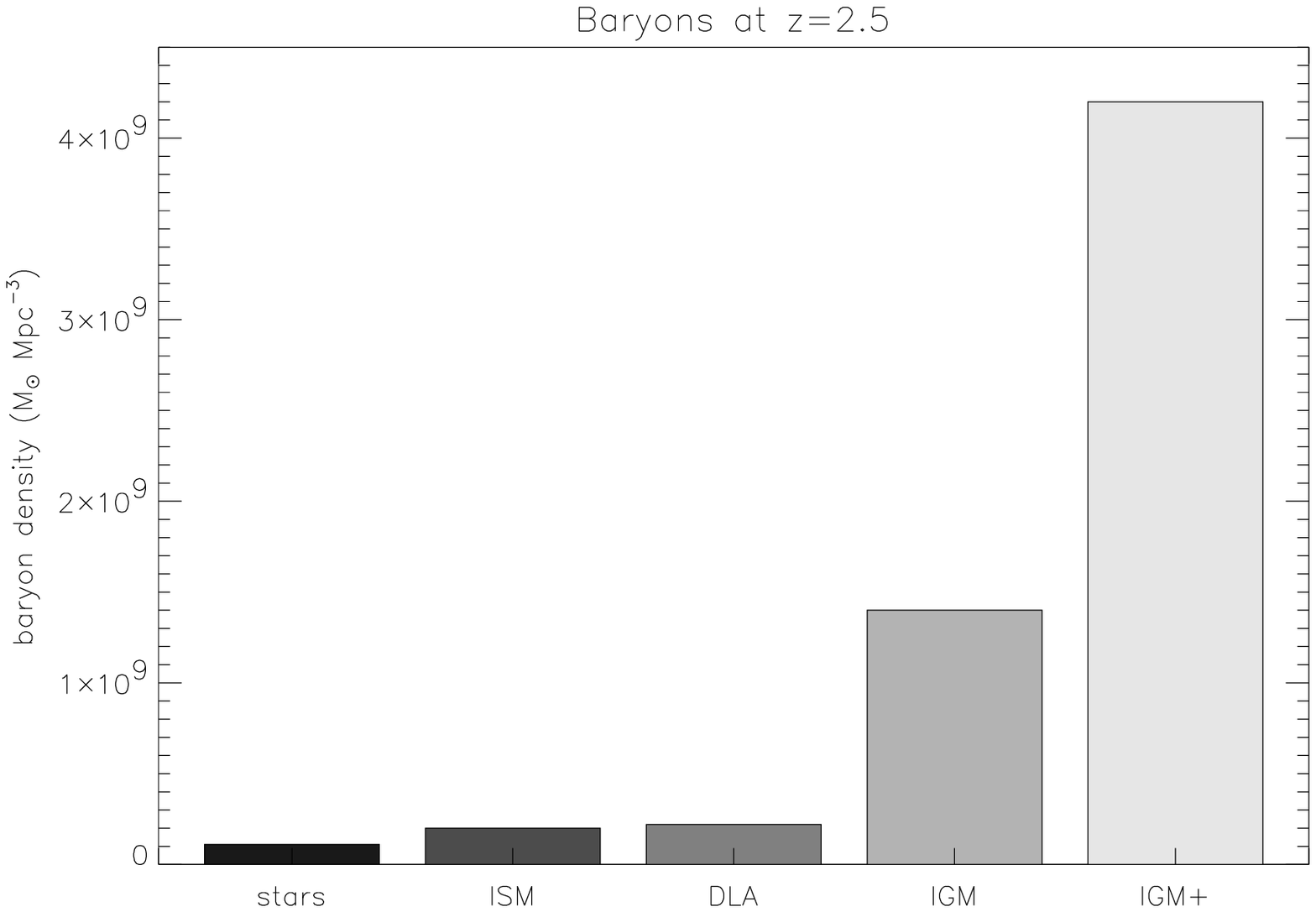,height=7.5cm,width=8.5cm}}\\
\caption{\label{histF} The location of metals and baryons
at low and high redshift. The categories are taken from
Table~\ref{metalsT}. DLAs are given as a separate category because it
is not clear whether they are produced by interstellar gas or
intergalactic gas, or a mixture of both. The darker shaded IGM
represents that currently observed -- the hot cluster gas at low
redshift and the Lyman forest clouds at high redshift. The lighter
IGM+ represents the range of values of extra IGM which have been
inferred for the group/field gas at low redshift and from CDM
predictions for the total Lyman forest gas at high
redshift (FHP). The unshaded regions for ISM and IGM+ in (a) and (c)
represent the maximum which could be present within the uncertainties
described in Table~\ref{metalsT}. For low-z ISM, this is when the
integral of the DMF is extrapolated to zero mass.
The additional IGM+ gas would appear to make up most of
the baryonic content at all redshifts (without it, there is a serious
shortfall in accounting for the baryon content expected from CMB
analysis and nucleosynthesis), but is not an important
location for metals at high redshift. Most metals at high-z are
evidently in the ISM, while at low redshift they are in stars and,
possibly, the low density IGM+.}
\end{figure*}

There is growing concern that the metal content of the high-z universe
is not observed to be as large as the inferred star formation would
suggest (Pettini 1999; Pagel 2002; Blain et al. 1999a,b). Accounting
for the metals observed in DLAs and the Lyman forest leaves a
shortfall of $\sim90$\% in the cosmic metal budget at $z=2-3$. Metals
are thought to be produced by supernovae and in the winds of cool
evolved stars. They are then injected into the ISM with a fraction of
them ($\eta$) bound into, or subsequently accreting onto dust
grains. From here they can be expelled out of the galaxy into the
inter-galactic medium (IGM) by starburst winds, or become part of the
next generation of stars. The fraction of metals found in these
different locations today is summarised in Fig.~\ref{histF}a. 

These fractions do not remain constant, of course, as the fraction of
total metals in stars and the IGM will decrease with redshift. Today
the fraction of metals in the ISM of galaxies is negligible, but how
different could this have been in the past, when galaxies were more
gas rich? It is clear that more metals should be found in the most
evolved objects at a given redshift, although their distribution
between stars and ISM will depend on how far the object has
evolved. This has been presented by Cen \& Ostriker (1999) as a
correlation between over-density and metallicity, the fastest evolving
objects being found preferentially in the most over-dense regions. The
monolithic collapse models predict massive episodes of star formation
at high redshift, and this should be accompanied by rapid enrichment
of the locality. Observations of DLAs and the IGM at high redshift
have failed to find evidence for this level of enrichment, although
high-z QSOs are often found to have near solar or super-solar
metallicities (Warner et al. 2002). This makes sense if QSOs represent
the accretion phase of the massive black hole, which is presumably
forming co-evally with the stellar bulge, as required to produce the
BH-bulge mass correlation in the local universe (Magorrian et
al. 1998). The IGM is a low-density environment and so not expected to
be very enriched at high-z. It has been argued that the Damped Lyman
alpha absorbers represent the co-moving mass density of gas in
galaxies at high redshift and that the column density-weighted
metallicity of these systems traces the global metallicity evolution
of the universe\footnote{provided that no class of galaxy is excluded
from the DLA samples} (Pei \& Fall 1995). However, there are potential
biases in the selection of DLAs as they are only observed along sight
lines to optically bright QSOs and thus they are less likely to be
observed if they contain dense gas and dust. The cross-sectional area
available at the edges of galaxies is also greater than the denser,
more evolved inner regions and therefore a sight-line is more likely
to intercept a less dense and less metal enriched part of a galaxy
(Mathlin et al. 2001). Thus current samples of DLAs may underestimate
the metal content of the neutral gas at any redshift. The obvious
place to look for the most enriched systems at high-z is to look at
the most obscured and actively star forming objects -- submm sources,
especially as they show evidence for being located in over-dense
regions (Ivison et al. 2000).

One simple thing we can do with the high-z DMF is to integrate the
total amount of dust in galaxies at low and high redshift. The
co-moving density of dust is given by $\rho_d =
\Gamma(2+\alpha)M^{\ast}_{\rm{d}} \phi^{\ast}$, where
$M^{\ast}_{\rm{d}}$ and $\phi^{\ast}$ are the Schechter parameters
fitted to the dust mass functions in Fig.~\ref{dmfF}. $\rho_d$ can be
converted into $\rho_z$, the co-moving density of metals, by dividing
by $\eta$ -- the fraction of ISM metals which are in dust. We have
used $\eta=0.4$ and there is some evidence that this fraction is
fairly constant both locally and at higher redshifts (Pei, Fall \&
Hauser 1999; Issa, MacLaren \& Wolfendale 1990; Edmunds 2001; James et
al. 2002). Table~\ref{metalsT} lists the ISM metals at low and high-z
inferred from the dust, along with an inventory of metals and baryons
in other locations. Integrating the DMF is currently quite uncertain
at low redshift due to the very steep faint end slope fitted to the
local DMF\footnote{This may simply be because we have not yet probed
to low enough dust masses to properly sample the turn over}. The
minimum and maximum values in Table~\ref{metalsT} are obtained by
integrating the local DMF down to the lowest observed mass, and to
zero mass respectively. To facilitate comparisons with work from the
literature, we now restrict ourselves to the $\Omega_m=1$ cosmology,
noting that this does not affect any of our conclusions.

\renewcommand{\baselinestretch}{1.0}
\begin{table*}
\centering
\caption{\label{metalsT}{Baryons and Metals in the
local and high redshift universe}}
\begin{tabular}{lccccc}
\\[-2.0ex] 
\hline
\\[-2.5ex]
\multicolumn{1}{l}{Location}&\multicolumn{2}{c}{Baryons
($\rm{M_{\odot}\,Mpc^{-3}}$)}&\multicolumn{1}{c}{}&\multicolumn{2}{c}{Metals ($\rm{M_{\odot}\,Mpc^{-3}}$)}\\
\multicolumn{1}{c}{}&\multicolumn{1}{c}{$z=0$}&\multicolumn{1}{c}{$z=2.5$}&\multicolumn{1}{c}{}&\multicolumn{1}{c}{$z=0$}&\multicolumn{1}{c}{$z=2.5$}\\
\hline
\\[-1.0ex]
stars & $5\times 10^{8} \, \rm{h_{75}}\,^{a}$ & $1.1\times 10^8
\,\rm{h_{75}}$ & &
$9.6\times 10^6 \, \rm{h_{75}}$ & $\approx 4.4\times10^{5}\,
\rm{h_{75}}$\\ 
\\[-2.5ex]
 &  & $\int$ SFH & & $Z=Z_{\odot}\,^b$ & see text \\
\\[-1ex] 
ISM  & $9.2\times 10^7 \, \rm{h_{75}\,^{a}}$ & $1.3 -
2.7\times 10^8 \, \rm{h_{75}}$ & & $0.8-3.0\times
10^5 \, \rm{h_{75}}$ & $1.9-3.8\times 10^6 \,\rm{h_{75}}$\\
\\[-2.5ex]
  & $\rm{H_{I}+H_2}$  & DMF \& model & & & $\langle \rm{T_d}\rangle = 25-20$ K\\
\\[-1ex]
DLAs & & $2.2\times 10^8 \, \rm{h_{75}\,^{a}}$ & & & $2.9\times 10^5
\,\rm{h_{75}}$\\
\\[-2.5ex]
 &  & ISM / IGM? & &  & $Z = 0.07Z_{\odot}\,^b$\\
\\[-1ex]
IGM observed & $3.7\times 10^8\, \rm{h_{75}^{0.5}\,^{a}}$ & 
$1.4\times 10^9 \,\rm{h_{75}\,^{a}}$ &  & $2.3\times 10^6
\,\rm{h_{75}^{0.5}}$ & $8.2\times 10^4 \,\rm{h_{75}}$\\
\\[-2.5ex]
 & cluster gas & Ly$\alpha$ forest & & $Z=0.33Z_{\odot}\,^d$ & $Z=0.003Z_{\odot}\,^b$\\
\\[-1ex]
IGM+ & $1-4.4\times 10^9\, \rm{h_{75}\,^{a}}$ &
$4.2\times 10^9 \,\rm{h_{75}\,^{a}}$  & & $1.9-8.3 \times 10^6
\,\rm{h_{75}}$ & $2.4\times 10^5 \,\rm{h_{75}}$\\
\\[-2.5ex]
  & group/field gas & extra inferred& & $Z=0.1\,Z_{\odot}$
& $Z=0.003Z_{\odot}\,^b$\\
 & & by CDM & & & \\
\\[-1ex]  
\hline
\\[-2.0ex]
Total observed & $1 - 5.4 \times 10^{9}\, \rm{h_{75}}$ &
$1.9-6.2\times 10^{9} \,\rm{h_{75}}$ & & $1.2-2.1\times 10^7 \,\rm{h_{75}}$ &
$2.7-4.9\times 10^6 \,\rm{h_{75}}$\\
\\[-1ex]
Inferred & $6.1\times 10^9\,^{c}$ & $6.1\times 10^9 \,^{c}$ & & $1-2\times
10^{7}$ & $2.5-5\times 10^{6}$\\
\\[-2ex] 
\\[-2.5ex] 
\hline 
\end{tabular}
\flushleft 
Table adapted from Pettini (1999) and Pagel (2002) with $\rm{h_{75} =
H_0/75\, km\, s^{-1}\, Mpc^{-1}}$ and $\Omega_{\rm{m}} =1$ for
consistency with previous work. We assume $Z_{\odot} = 0.0189$. The
metal content has been estimated from either the baryon content and
the metallicity, or from the dust masses. The `IGM observed'
represents the directly observed gas (the hot cluster gas at low-z,
and the Ly$\alpha$ clouds at high-z). The `IGM+' is the additional IGM
gas which has been inferred but not directly observed (the group/field
gas locally, and predictions for the total IGM from CDM simulations at
high-z). The `total observed' row has a minimum when counting only
`IGM observed' and a maximum when `IGM+' is included. All values are
from the literature except the following: The metal content of the ISM
at both high and low-z, and the baryon content of the ISM at high-z,
and the stellar metal content at high-z have been estimated as
described in the text. The metallicity of the postulated group/field
gas at low-z has been estimated to be $\sim 0.1\, Z_{\odot}$ but this
has not yet been measured. The bottom line shows the baryon density
inferred from the CMB measurements (Netterfield et al. 2002), and the
metal content of the universe at high and low-z predicted from the
integrated star formation history (Pettini 1999; Pagel 2002), with a
yield of 0.021--0.042.

Refs: $^a$ Fukugita, Hogan \& Peebles (1998). $^b$ Pettini
(1999). $^c$ Netterfield et al. (2002). $^d$ Pagel (2002)

\end{table*}

\begin{figure}
\psfig{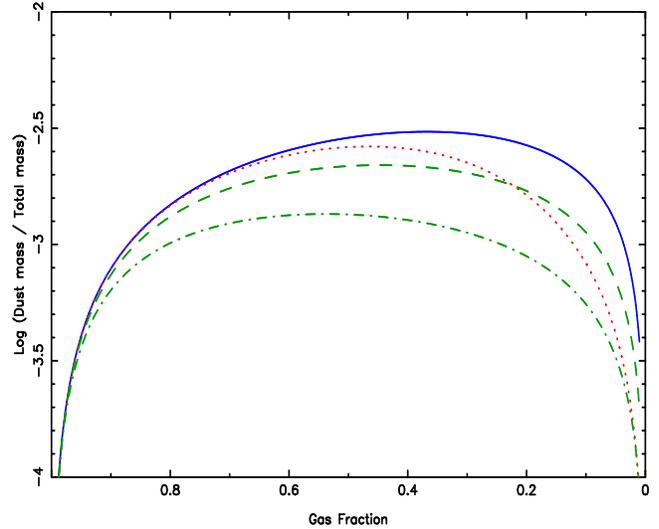}
\caption{\label{maxdmF} Models of how the dust mass of a galaxy
changes as it evolves. The y-axis shows the predicted dust mass
divided by the total baryonic mass of the galaxy. The x-axis shows the
fraction of the baryonic mass that is in the form of gas; as the
galaxy evolves, it should move from left to right. Spiral galaxies
today have a gas fraction of 0.1--0.2, elliptical galaxies have a gas
fraction closer to zero, but these simple models break down at such
low gas fractions and can not give accurate predictions for the dust
to baryonic mass ratio of elliptical galaxies today. The models are a
closed-box (solid) in which no gas is either lost or accreted by the
galaxy, a limiting inflow model in which gas is accreted (dotted), and
two outflow models, where gas is lost at a rate proportional to either
1 or 4 times the star formation rate (dashed, dot-dashed). Further
details are in Eales \& Edmunds (1996). The models require assumptions
about the fraction of metals bound up in dust grains ($\eta=0.4$) and
the global yield ($p=0.021$).}
\end{figure}

The baryon values are taken from the literature (Fukugita, Hogan \&
Peebles 1998 -- FHP) except for that associated with the ISM at
high-z. For this we have used the metals in dust, plus a simple
`closed-box' chemical evolution model to predict the baryon content of
the ISM at high redshift. This chemical evolution model
(Fig.~\ref{maxdmF}), and others which incorporate inflow and outflow
of gas, predict that the ratio of dust mass to baryonic mass should
change quite slowly over a large part of a galaxy's evolution, and
should peak when roughly half the gas has been converted into stars
(Eales \& Edmunds 1996, Edmunds \& Eales 1998). We have used the
maximum ratio of dust mass to baryonic mass (we are assuming
that the submm sources are observed at their maximum dust mass) given by the
model, and the integral of the DMF, to estimate the total amount of
baryons associated with the objects in the DMF. Since the maximum
occurs at a gas fraction of $\sim 0.5$, the
total baryonic mass is split equally between stars and gas (ISM). The
assumption that the submm sources are observed close to their peak
dust mass is probably valid for the bright submm selected sources, but
cannot be presumed to hold for the faintest lensed sources and the
statistical detection of optically selected sources in the HDF. This
means that the baryon density we have associated with the ISM in
Table~\ref{metalsT} is really a minimum.

The results in Table~\ref{metalsT} and Fig.~\ref{histF} show a number
of different things. First, there was $6-45$ times more dust in
galaxies at $z=2.5$ than there is today (assuming the dust properties
used to estimate the masses are similar at low and high-z, most of the
uncertainty here comes from the integration of the local
DMF). Secondly, as illustrated in Fig.~\ref{histF}, the amount of
metals traced by dust at high-z is 3--10 times the total amount of
metals in DLA and Lyman forest systems. The Lyman forest represents
the IGM at high-z, while the role of the DLAs is less clear. They may
be the ISM of galaxies (Wolfe et al. 1986), either the outer parts of
galactic disks or their precursors (Efstathiou 2000; Boissier, Peroux
\& Pettini 2002), or dwarf galaxies (Yanny
\& York 1992). Some fraction of them may also be linked to
outflows from star-forming galaxies and therefore could be classed as
IGM (Shaye 2001). Whichever is the case, at
high-z the metal content of the ISM is clearly much greater than that
of the IGM. At low redshift the situation is reversed: the IGM now
contains far more metals than the ISM of galaxies. This is due to two
processes: the consumption of ISM gas and metals by star formation and
the expulsion of metal enriched gas from galaxies by winds. Much of
the IGM today may be contained in groups of galaxies (this is shown
in Fig.~\ref{histF} and Table~\ref{metalsT} as `IGM+'). The amount of this
gas is very uncertain but could be enough to provide the baryonic
matter necessary to match that inferred from recent CMB measurements
(Netterfield et al. 2002). The amount of metals which may be contained
in the postulated group IGM gas is even more speculative. The hot IGM gas in
rich clusters has a metallicity of $\sim 0.3 \, Z_{\odot}$ (Renzini 1997)
but the metallicity of gas in less dense environments, such as in
groups, may be lower (as predicted by Cen \& Ostriker 1999). We have
used an estimate of $Z=0.1\,Z_{\odot}$ for the `IGM+' metallicity but
stress the large uncertainty in both the amount of gas and metals in
this category.

We have also included in Table~\ref{metalsT} an admittedly uncertain
estimate of the total amount of metals contained in stars at $z\sim
2.5$. This was estimated from the following three populations: 
\begin{enumerate}
\item{Lyman Break Galaxies (LBG) -- we took the $z=3$ LF from Shapley
et al. (2001), corrected to this cosmology ($\Omega_m=1$), and
integrated to $0.2 L^{\ast}$. Using $\rm{(B-V)=0.16}$ and $M/L_B=0.15$
(Pettini et al. 2001) we used the integral of the LF to estimate the
stellar mass density $\rho_{\ast} = 4\times 10^7\, \rm{M_{\odot}
Mpc^{-3}\, h_{75}}$. We used a metallicity of $Z=0.1-0.3
Z_{\odot}$ (Pettini et al. 2001) to produce an estimate of the metal
density in stars in bright LBG: $\rho_{z,\ast}(\rm{LBG}) \sim 7.5\times 10^4 -
2.3\times 10^5 \,\rm{M_{\odot}\, Mpc^{-3}\, h_{75}}$. An obvious uncertainty
here is the correction to the luminosity density required for dust
extinction, which could be a factor 2--5}
\item{Submm selected sources ($>3$ mJy) -- the dust mass density (from
integrating the dust mass function down to 3 mJy), and the closed box
model were used to estimate the mass of metals in stars (and the mass
in stars). As above, we assumed that the galaxies were halfway through
their evolution, with a gas fraction of $\sim 0.5$. This gave a
stellar mass density of $\rho_{\ast} = 3.8\times 10^7
\,\rm{M_{\odot}\, Mpc^{-3}\, h_{75}}$ and a stellar metal density of
$\rho_{z,\ast}(\rm{sub}) \sim 2.4\times 10^5\, \rm{M_{\odot}\,
Mpc^{-3}\, h_{75}}$. The main uncertainties here are the temperature
of the dust (estimates will increase by a factor 2 if the temperature
is 20 K rather than 25 K), and the broadness of the maximum in the
ratio of dust mass/total mass (Fig.~\ref{maxdmF}). This means there is
a considerable range in the gas fraction and hence the stellar mass
associated with a given dust mass. Using a model with inflow or
outflow will lead to one inferring a greater baryonic mass from a
given dust mass}
\item{The rest of the stellar mass at $z=2.5$ predicted from the SFH and
not in bright LBG or SCUBA sources. We assumed that this is in lower
luminosity systems and DLAs and has $Z=0.07\, Z_{\odot}$ (the average
DLA metallicity (Pettini 1999)).}
\end{enumerate}
The sum of these three populations is designed to be equivalent to the
total stellar density predicted from the SFH, and has a mean metallicity
of $\sim 0.2\, Z_{\odot}$. Correction of the LBG values for dust would
mean less stars in category (iii) and more stellar metals overall (as
LBG have higher metallicities). A comparison of the metals traced by
dust (ISM) and those in stars at the two epochs leads to a third
conclusion: at low redshift the ratio of metals in the ISM to those in
stars is only $\leq 0.08$, while at $z\sim 2.5$ this ratio is $\sim
4-9$. The chemical evolution model suggests that the ratio will be as
high as this when only $\sim 20-30$\% of stars in the universe have
formed, in agreement with the star formation history inferred from
observations. Taken together, this means that most metals at $z=2-3$
are in the ISM of galaxies. This is as expected in the early stages of
galaxy evolution; later as the gas is consumed, the metals become
locked into stars and expelled into the IGM.

Table~\ref{metalsT} shows that we can now account for all the metals
predicted to exist (from integrating the SFH) at both low and high
redshift, and thus there is no longer a `missing metals' problem. The
DLAs do not seem to trace the bulk of the metals within galaxies at
high-z, suggesting that there are indeed biases present in DLA samples
against regions of high density and obscuration (Pei \& Fall 1995). A
survey of DLA's selected in the radio (Ellison et al. 2001) has found
that there is no evidence for a large missing population of high
column density absorbers. The data allow for $\Omega_{\rm{DLA}}$ to
increase by a factor of 2 at most. However, the statistics of the
survey are still somewhat limited, especially at the highest column
densities where one may also expect to find the greatest quantities of
metals (in an unbiased survey). We have
also shown that the `missing metals' can be accounted for without
resorting to a large reservoir of metals in a hot, diffuse and as yet
undetected part of the IGM at high redshift, as has been suggested by
various authors (Pettini 1999; Pagel 2002). However,
there is still a `missing baryons' problem, since the total amount of
directly observed baryons is less than the amount inferred from recent
microwave background (CMB) measurements (Netterfield et
al. 2002). This shortfall is currently thought to be made up of
inter-galactic gas at both low and high redshift (FHP) (`IGM+' in
Table~\ref{metalsT} and Fig.~\ref{histF}). 

\section{Dust mass evolution and the spheroid connection.}

\begin{figure}
\psfig{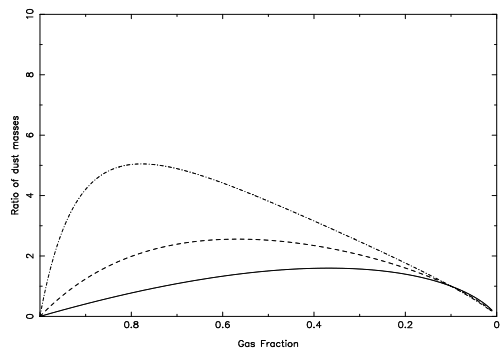}
\caption{\label{modelF} Predicted dust mass for a spiral galaxy versus gas
fraction. In this figure, the dust mass has been divided by the dust
mass it has at a gas fraction of 0.1. The
curves are the predictions of the closed-box and outflow models
described in the caption of Fig.~\ref{maxdmF}.}
\end{figure}

The chemical evolution models can also be used to try to explain the
evolution in the dust mass function shown in
Fig.~\ref{dmfF}. Fig.~\ref{modelF} shows how the ratio of dust mass at
any point in the evolution of a spiral galaxy compares with that
today. Present-day spiral galaxies have a gas fraction of $\sim
0.1-0.2$, and so in Fig.~\ref{modelF} we have shown how the mass of
dust in a galaxy is predicted to change relative to the dust mass it
has at $f=0.1$. The figure shows that there is relatively little scope
for the evolution of the dust masses of spirals; they can have at the
most dust masses which are factor of 4 greater in the past than their
dust masses today. The local DMF is composed entirely of spiral
galaxies, and using pure luminosity evolution (or, in this case, dust
mass evolution) requires an increase in the dust masses of the local
galaxies of a factor $\sim 10$ in order to reproduce the high redshift
objects with $M_{\rm{d}}>M_{\rm{d}}^{\ast}$. Due to the different
shapes of the high and low redshift DMFs, the discrepancy is not as
large for the lowest mass objects, and so the dust masses of the
faintest submm sources (from the statistical analysis of optically
selected starbursts in the HDF) are consistent with simple evolution
from the local DMF. Thus the high redshift DMF cannot be reproduced by a
chemical evolution model applied to local spiral galaxies, which argues that
the high redshift dust sources found in the deep SCUBA surveys are a
population which is not represented in the dust mass function today,
i.e. that the high redshift sources are in fact the ancestors of the
gas and dust poor elliptical galaxies in the local universe.

Present-day elliptical galaxies have a much lower typical gas fraction
$\sim 0.01$ and if Fig.~\ref{modelF} were to be drawn for ellipticals
(showing how dust mass in the past changes relative to that at a gas
fraction of 0.01), we would see curves of similar shape to those in
Fig.~\ref{modelF} but which peak at a maximum of between 10 and 35
times more dust in the past. However, we note that because the gas
fractions and dust masses of present-day ellipticals are poorly known,
and because the ratio in Fig.~\ref{modelF} becomes very sensitive to
$f$ at low gas fractions\footnote{For example, if the gas fraction
today were 0.001 then the maximum increases in dust mass allowed
become 50 -- 250.} these simple models are unable to predict
accurately the factor by which the dust mass of ellipticals could have
been greater at their peak dust content. By the time elliptical
galaxies have reached their current evolutionary status, there will be
competition between grain formation in low-mass giant stellar
atmospheres ($\sim 0.8-1 \,
\rm{M_{\odot}}$) and the very occasional SNae-Ia, and long-term
erosion of grains by hot X-ray gas. These are very different
conditions from the grain formation in massive and intermediate-mass
stellar envelopes, SNae-II, mantle growth in interstellar clouds and
ISM gas conditions that were present at the peak of
their dust content. Until a comprehensive model can be based on
reliable measurements of the gas and dust content of present-day
ellipticals, it is probably unwise to try to extrapolate the dust
masses ellipticals should have possessed at their peak dust content
(robust value) to the dust masses they should have today (very
uncertain), other than to accept
that there are good reasons why the dust masses could have been much
greater in the past (by factors 10--100).

The highest ratio in Fig.~\ref{modelF} is for an outflow
proportional to 4 times the star formation rate. This is quite
extreme, and leads to an original mass 5 times greater than the
remnant today, which would be difficult to produce in most
hierarchical models of galaxy formation. Outflow episodes are expected
in a galaxy's lifetime but at a much lower rate, or for a shorter
time. Table~\ref{metalsT} also shows that the change in ISM baryons
from high redshift to today is of the same order as the increase in
stellar baryons. This means there is not scope for 70--80\% of the ISM
to have outflowed between $z\sim 2.5$ and today. Some level of outflow
is probably necessary to produce sufficient evolution in the dust
masses of ellipticals, and to provide an obvious source of at least
some of the inter-galactic metals observed in the Lyman-forest at high
redshift, and the inter-cluster medium today. However, the elementary
outflow models can give rise to problems in interpreting the
nucleosynthetic yields and metallicities.

We now use the results so far to address the hypothesis that SCUBA
galaxies are the ancestors of today's massive ($L>L^{\ast}$)
spheroids. First, we compare the baryons contained in the two
populations. At low-z, we construct a $b_j$ optical spheroid luminosity
function (LF) from the 2dF redshift survey (Folkes et
al. 1999) and type specific bulge/disk ratios (FHP). We then multiply
by the stellar mass-to-light ratio (FHP)
$(M/L)_{\ast}= 4.5-8.3$ to give a spheroid baryonic mass function
(assuming that in today's spheroids, the stars dominate the baryonic
mass). This is integrated down to $M^{\ast}_b$ to give the mass density
in $L>L^{\ast}$ spheroids as 
$\rm{\rho_{sph,b}} = 8.5-15.6\times 10^7\,
\rm{M_{\odot}\, Mpc^{-3}\, h_{75}}$. At high-z, we integrate the DMF
down to a mass equivalent to an 850$\mu$m flux of 3 mJy ($\sim
M^{\ast}_d$). We then use the ratio of $\rm{M_{dust}/M_{baryon}}$ from
the closed box model (Fig.~\ref{maxdmF}) at a gas fraction of 0.5 to
estimate the baryon content of the submm sources, $\rm{\rho_{sub,b}} =
7.6 - 15.3\times 10^7\,
\rm{M_{\odot}\, Mpc^{-3}\, h_{75}}$. Since we use the ratio at its
maximum (which is very broad) we are being conservative in our
estimate; using outflow or inflow would also increase the inferred
$\rm{M_{baryon}}$. The baryonic content of the SCUBA sources at high-z
is therefore similar to the baryonic content of today's massive
spheroids, the distant submm sources simply being observed at a time
when a greater fraction of those baryons were in the form of gas
rather than stars. There is a further interesting result, namely that
if we compare the baryons at $z=2.5$ associated with the SCUBA sources
to those in DLAs, we get the same ratio (1.2 -- 2.4) as when we
compare the baryons in $L>L^{\ast}$ spheroids and all disk
galaxies today (FHP). This coincidence supports the idea that spiral
disks are the descendants of the high-z DLA population (Wolfe et
al. 1986; Pettini 1999), and that
the spheroids are the descendants of the SCUBA sources. That there are
far more metals in the SCUBA sources than the DLAs is evidence that
the bulk of stars in spheroids formed at earlier times than those in
disks. 

We can also use the chemical evolution model to predict the dust mass
function that should have been produced by the spheroids when they had
formed $\sim 50$\% of their stars. We again take the local spheroid LF
and multiply by the stellar mass/light ratio to give a spheroid mass
function. Using the closed-box model from Fig.~\ref{maxdmF}, we then
multiply the spheroid masses by the maximum ratio of dust/baryonic
mass to give a dust mass function. This is shown as the shaded region
in Fig.~\ref{dmfF}. Including outflow has little effect on this shaded
region, as the decrease in the maximum dust/baryonic mass ratio
(Fig.~\ref{maxdmF}) is
offset by the increase in overall baryonic mass at higher gas
fractions. The agreement between this very simple prediction and the
observed DMF is rather good, supporting the idea that the descendants
of SCUBA galaxies are not the dusty objects in our low-z DMF, but the
now dust-poor spheroids. Putting this another way, bright submm
selected sources ($>3$ mJy) can account for all the predicted dust
from $L>L^{\ast}$ spheroids.
 
\section{Conclusions}

We have used submillimetre surveys of the local and distant universe
to produce dust mass functions. Taken together these show the
evolution in dust mass and ISM metals from redshift zero to
$z\sim2.5$. In conjunction with a simple chemical evolution model we
have found that:

\begin{enumerate}
\item{There was much more dust in galaxies in the past. Simple dust
mass evolution in local dusty galaxies to match the high-z dust mass
function is not feasible as the chemical models do not permit spiral
galaxies to have had $\times 10$ more dust at an earlier point in
their history. This argues that the local submm sources are not the
descendants of the sources in the deep surveys.}
\item{Most metals ($>70$\%) at $z\sim 2.5$ are found in the ISM of
galaxies -- preferentially in the most obscured and active
objects. This solves the `missing metals' problem, which was evident
when comparing the metals observed in DLAs and the Lyman forest with the
metal density inferred from integrating the star formation
history. This supports the suggestion (Pei \& Fall 1995) that DLA
samples are not unbiased tracers of star forming galaxies at high-z,
but instead sample the lower density/unobscured (and therefore
un-enriched) environments.}
\item{Comparison of baryons in $\rm{L>L^{\ast}}$ spheroids today, and
$\rm{M_d>M_d^{\ast}}$ submm sources at high-z, show good agreement, indicating
that the submm sources are plausibly the ancestors of the
spheroids. Using the local optical spheroid LF and the chemical
evolution model, we can also predict the dust mass function the
spheroids should have had at the time when they had formed $\sim 50$\%
of their stars. This agrees remarkably well with the observed dust
mass function at high-z, and again strongly supports the hypothesis
that SCUBA sources (particularly brighter ones) are elliptical
galaxies in the process of forming most of their stars.}
\end{enumerate}


We would like to thank the referee for helpful suggestions on
improving the manuscript and also Jon Davies, Sarah Ellison and
Bernard Pagel for useful comments. L. Dunne aknowledges the support
of a PPARC fellowship. S. Eales thanks the Leverhulme Trust for the
award of a research fellowship.
{}


\begin{thebibliography}{}


\bibitem{} Alton P. B., Bianchi S., Rand R. J., Xilouris E., Davies
J. I., Trewhella M., 1998, ApJL, 507, L125

\bibitem{} Alton P. B., Lequeux J., Bianchi S.,
 Churches D., Davies, J. Combes F., 2001, A\&A, 366, 451

\bibitem{} Bianchi S., Davies J. I. \& Alton P. B., 1999, 344 L1.

\bibitem{} Blain A. W., Smail I., Ivison R. J., Kneib J.-P., 1999a,
MNRAS, 302, 632

\bibitem{} Blain A. W., Jameson A., Smail I., Longair M. S., Kneib
J.P., Ivison R. J., 1999b, MNRAS, 309, 715

\bibitem{} Boissier S., P\'{e}roux C., Pettini M., 2002, MNRAS
accepted. (astro-ph/0208457)

\bibitem{} Borys C., Chapman S. C., Halpern M., Scott D., 2002,
MNRAS, 330, L63. 

\bibitem{} Bower R. G., Lucey J. R. \& Ellis R. S., 1992, MNRAS,
254, 601.

\bibitem{} Braine J., Gu\'{e}lin M., Dumke M., Brouillet N., Herpin
F., Wielebinski R., 1997, A\&A, 326, 963.

\bibitem{} Calzetti D., Armus L., Koorneef J., Storchi-Bergmann T.,
2000, ApJ, 533, 682.

\bibitem {} Carilli C. L., Yun M. S., 1999, ApJ, 513, L13

\bibitem{} Cen R. \& Ostriker J. P., 1999, ApJ, 519, L107.

\bibitem{} Chapman S. C., Smail I., Ivison R. J., Helou G., Dale
D. A., Lagache G., 2002, ApJ, 573, 66.


\bibitem{} Draine B. T., Lee H. M., 1984, ApJ, 285, 89

\bibitem{} Dumke M., Braine J., Krause M., Zylka R., Wielebinski R.,
Gu\'{e}lin M., 1997, A\&A, 325, 124

\bibitem{} Dunlop, J. S., 2001a,  New Astronomy Reviews, 45, 609.

\bibitem{} Dunlop, J. S., 2001b, in Lowenthal J., Hughes D. eds, Deep
Millimetre Surveys, World Scientific (Singapore), p. 11

\bibitem{} Dunne L. et al., 2000, MNRAS, 315, 115. 

\bibitem{} Dunne L., Clements D. L. \& Eales S. A., 2000, MNRAS,
319, 813. 

\bibitem{} Dunne L. \& Eales S. A., 2001, MNRAS, 327, 697.

\bibitem{} Eales S. A. \& Edmunds M. G., 1996, MNRAS, 280, 1167.

\bibitem {} Eales S. A., Lilly S. J., Gear W. K. P., Dunne L., Bond
J. R., Hammer F., Le F\`{e}vre O., Crampton D., 1999, ApJ, 515, 518

\bibitem{} Eales S. A. et al., 2000, AJ, 120, 2244. 


\bibitem{} Edmunds M. G., Eales S. A., 1998, MNRAS, 299, L29

\bibitem{} Edmunds M. G., 2001, MNRAS, 328, 223.

\bibitem{} Efstathiou G., 2000, MNRAS, 317, 697

\bibitem{} Ellison S. L., Yan L., Hook I. M., Pettini M., Wall J. V.,
Shaver P., 2001, A\&A, 379, 393.


\bibitem{} Folkes S. et al., 1999, MNRAS, 308, 459.

\bibitem{} Frayer D. T., Ivison R. J., Smail I., Yun M. S., Armus L.,
1999, AJ, 118, 139

\bibitem{} Frayer D. T., 2001, in Lowenthal J., Hughes D. eds, Deep
Millimetre Surveys, World Scientific (Singapore), p. 117

\bibitem{} Fukugita M., Hogan C. J. \& Peebles P. J. E., 1998, ApJ,
503, 518. (FHP)

\bibitem{} Granato G. L., Silva L., Monaco P., Panuzzo P., Salucci P.,
de Zotti G., Danese L., 2001, MNRAS, 324, 757.

\bibitem{} Gu\'{e}lin M., Zylka R., Mezger P. G., Haslam C. T., Kreysa 
E., 1995, A\&A, 298, L29

\bibitem{} Haas M., 1998, A\&A, 337, L1

\bibitem{} Haas M., Klaas U., Coulson I., Thommes E., Xu C., 2000,
A\&A, 356, L83

\bibitem{} Hildebrand R. H., 1983, Q. J. R. Astron. Soc., 24, 267.


\bibitem {} Hughes D. H. et al., 1998, Nature, 394, 241.

\bibitem{} Isaak K. G. et al., 2002, MNRAS, 329, 149.

\bibitem{} Issa M. R., MacLaren I. \& Wolfendale A. W., 1990, A\&A,
236, 237.

\bibitem{} Ivison R. J., Dunlop J. S., Smail I., Dey A., Liu M. C.,
Graham J. R., 2000, ApJ, 542, 27.

\bibitem{} Ivison R. J., Smail I., Frayer D. T., Kneib J.-P., Blain A. W., 2001, ApJ, 561, L45

\bibitem{} Ivison R. J. et al., 2002, MNRAS in press. (astro-ph/0206432).

\bibitem{} James A., Dunne L., Eales S. A., Edmunds M. G., 2002,
MNRAS, 335, 753

\bibitem{} Jimenez R. et al., 1999, MNRAS, 305, L16.

\bibitem{} Klaas U. et al., 2001, A\&A, 379, 823.

\bibitem{} Kr\"{u}gel E., Steppe H., Chini R., 1990, A\&A, 229, 17

\bibitem{} Lilly S. J., Le Fevre O., Hammer F., Crampton D., 1996,
ApJ, 460, L1. 

\bibitem{} Lilly S. J. et al., 1999, ApJ, 518, 641.

\bibitem{} Lutz D. et al., 2001, A\&A, 378, 70

\bibitem{} Madau P., et al. 1996, MNRAS, 283, 1388

\bibitem{} Madau P., Pozzetti L., Dickinson M., 1998, ApJ, 498, 106

\bibitem{} Magorrian J. et al., 1998, AJ, 115, 2285

\bibitem{} Mathlin G. P., Baker A. C., Churches D. K., Edmunds M. G.,
2001, MNRAS, 321, 743

\bibitem{} Mihos J. C., 1999, AP\&SS, 266, 195

\bibitem{} Mihos J. C., Hernquist L., 1996, ApJ, 464, 641

\bibitem{} Neininger N., Gu\'{e}lin M., Garc\'{i}a-Burillo S., Zylka
R., Wielebinski R., 1996, A\&A, 310, 725

\bibitem{} Netterfield C. B. et al., 2002, ApJ, 571, 604.

\bibitem{} Pagel, B. E. J., 2002, in Matteucci, F. \& Giovannelli,
F. eds, Chemical Enrichment of Intracluster and Intergalactic Medium,
ASP Conf. Proc. Vol. 253, p. 489.

\bibitem{} Papadopoulos P. P., Seaquist E. R., 1999, ApJ, 514, L95

\bibitem{} Peacock J. A. et al., 2000, MNRAS, 318, 535. 

\bibitem{} Pei Y. C., Fall M. S., 1995, ApJ, 454, 69

\bibitem{} Pei Y. C., Fall M. S. \& Hauser M. G., 1999, ApJ, 522, 604.

\bibitem{} Pettini M., 1999, in Walsh, J. R. \& Rosa, M. R. eds, Chemical
Evolution from Zero to High Redshift. Springer-Verlag (Berlin) p. 233.

\bibitem{} Pettini M. et al., 2001, ApJ, 554, 981.

\bibitem{} Reach W. T. et al., 1995, ApJ, 451, 188

\bibitem{} Renzini A., 1997, ApJ, 488, 35 



\bibitem{} Schaye J., 2001, ApJ, 559, L1

\bibitem{} Scott S. E. et al., 2002, MNRAS, 331, 817.

\bibitem{} Shapley A. et al., 2001, ApJ, 562, 95.

\bibitem{} Sievers A. W., Reuter H.-P., Haslam C. T., Kreysa E., Lemke 
R., 1994, A\&A, 281, 681

\bibitem {} Smail I., Ivison R. J., Blain A. W., 1997, ApJ, 490, L5

\bibitem{} Smail, I., Ivison R. J., Blain A. W., Kneib J.-P., 2002,
MNRAS, 331, 495

\bibitem{} Sodroski T. J., Odegard N., Arendt R. G., Dwek E., Weiland
J. L., Hauser M. G., Kelsall T., 1997, ApJ, 480, 173

\bibitem{} Soifer B. T., Boehmer L., Neugebauer G., Sanders D. B.,
1989, AJ, 98, 766

\bibitem{} Steidel, C. C., Adelberger, K. L., Giavalisco, M., Dickinson,
M., Pettini, M., 1999, ApJ, 519, 1.

\bibitem{} Tissera P. B., Dom\'{i}nguez-Tenreiro R., Scannapieco C.,
S\'{a}iz A., 2002, MNRAS, 333, 327

\bibitem{} Trewhella M., Davies J. I., Alton P. B., Bianchi S.,
 Madore B. F., 2000, ApJ, 543, 153

\bibitem{} Warner C., Hamann F., Shields J. C., Constantin A., Foltz
C. B., Chaffee F. H., 2002, ApJ, 567, 68

\bibitem{} Webb T. M. A. et al., 2002, ApJ submitted. (astro-ph/0201180).

\bibitem{} Wolfe A. M., Turnshek D. A., Smith H. E., Cohen R. D.,
1986, ApJS, 61, 249

\bibitem{} Yanny B., York D. G., 1992, ApJ, 391, 569








































 






\end{thebibliography}
\end{document}